# Large magnetoelectric effect in mechanically mediated structure of TbFe$_2$, Pb(Zr,Ti)O$_3$, and nonmagnetic flakes

K. Bi,[1] Y. G. Wang,[1,a)] D. A. Pan,[2] and W. Wu[1]
[1]*College of Materials Science and Technology, Nanjing University of Aeronautics and Astronautics, Nanjing 210016, People's Republic of China*
[2]*School of Materials Science and Engineering, University of Science and Technology Beijing, Beijing 100083, People's Republic of China*

Magnetoelectric (ME) effect has been studied in a structure of a magnetostrictive TbFe$_2$ alloy, two piezoelectric Pb(Zr,Ti)O$_3$ (PZT) ceramics, and two nonmagnetic flakes. The ME coupling originates from the magnetic-mechanical-electric transform of the magnetostrictive effect in TbFe$_2$ and the piezoelectric effect in PZT by end bonding, instead of interface bonding. Large ME coefficients of 10.5 and 9.9 V cm$^{-1}$ Oe$^{-1}$ were obtained at the first planar acoustic and third bending resonance frequencies, which are larger than that of conventional layered TbFe$_2$/PZT composites. The results show that the large ME coupling can be achieved without interface coupling. © 2011 American Institute of Physics. [doi:10.1063/1.3574004]

Magnetoelectric (ME) materials have stimulated tremendous fundamental and practical interests due to their potential applications in the fields of sensors, actuators, and transducers.[1–4] Hereinto, layered ME composites exhibit excellent ME effect at room temperature compared to single phase materials and multiphase bulk composites.[5–9] The ME effect in layered ME composites is a product property of the magnetostrictive and piezoelectric effects. For most investigations on layered ME composites, the classical structure is laminated with layers by interface bonding. A magnetic field applied to the layered ME composites will induce strain in the magnetostrictive layer which is passed along to the piezoelectric layer by interface coupling, where it induces an electric polarization.[10] Therefore, the interface coupling is a key factor that determines the ME coupling of layered composites.[11] However, in practice, there is no ideal interface for layered ME composites, which reduces the ME coupling.

Large ME effect is not limited to be produced by interface coupling between magnetostrictive and piezoelectric phases, but will rather appear in the case when a magnetic force can mechanically act on a piezoelectric phase. Recently, the extrinsic ME effect was realized in multiphase magnet-piezoelectric composites which is not influenced by interface condition.[12–15] In this work, we report a large ME coupling obtained from a structure mechanically mediated by a stretching force and made up of a magnetostrictive TbFe$_2$ flake, two piezoelectric Pb(Zr,Ti)O$_3$ (PZT) flakes, and two nonmagnetic flakes. The proposed structure features an improved stretching stress coupling between the TbFe$_2$ and PZT flakes. And it is simple and more similar to the classical one than the multiphase magnet-piezoelectric composites,[12–15] which is advantageous for practical applications.

The proposed ME structure, as shown in Fig. 1(a), is made up of a magnetostrictive TbFe$_2$ flake, two piezoelectric PZT flakes, and two nonmagnetic (glass) flakes. The TbFe$_2$ flake was prepared by arc melting under argon atmosphere with dimensions of $15 \times 6 \times 3.5$ mm$^3$ ($l \times w \times t$). The commercial PZT flakes were sliced with dimensions of $15 \times 6 \times 0.6$ mm$^3$ and polarized along the thickness direction. Their dielectric, piezoelectric, and electromechanical coupling coefficients are $\varepsilon_{33}$=2400, $d_{33}$=400 pC N$^{-1}$, and $k_{33}$=0.70, respectively. Both TbFe$_2$ and PZT flakes were arranged parallel to the length direction. The glass slides, which is nonmagnetic or nonpiezoelectric, were bonded with both ends of the TbFe$_2$ and PZT flakes by epoxy and arranged perpendicular to the length direction.

The working principle of the proposed structure is essentially based on the stress-mediated product of the magnetostrictive effect in the TbFe$_2$ flake and the mechanically trans-

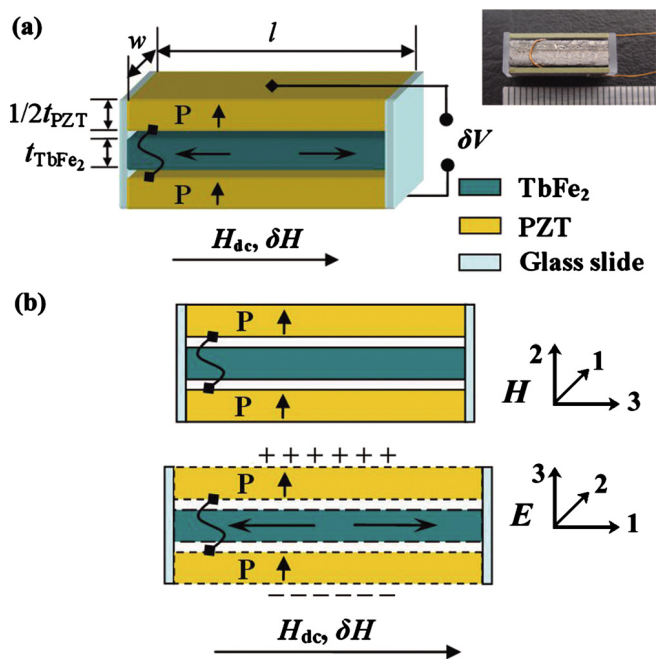

FIG. 1. (Color online) (a) Schematic diagram and photograph of the proposed structure of a magnetostrictive TbFe$_2$ flake, two piezoelectric PZT flakes, and two nonmagnetic flakes and (b) illustrations of the working principle.

a)Author to whom correspondence should be addressed. Electronic addresses: yingang.wang@nuaa.edu.cn and bike@nuaa.edu.cn.

formed piezoelectric effect in the PZT flake under end bonding. When an external ac magnetic field ($H_{ac}$) is applied along the length direction of magnetostrictive (TbFe$_2$) flake, it will produce a longitudinal shift as a result of the magnetostrictive effect. Hence, a stretching stress is produced in PZT flakes as the end parts of TbFe$_2$ and PZT flakes are bonded with glass slides in a vertical plane. It is spread throughout the ends. Due to the piezoelectric effect, the PZT flakes will induce a charge output, as illustrated in Fig. 1(b). In this case, the end bonding, instead of interface bonding, between magnetostrictive and piezoelectric phases can achieve ME coupling.

The piezomagnetic constitutive equations can be expressed as

$$S_{3m} = s_{33}^H T_{3m} + d_{33,m} H_3, \quad (1a)$$

$$B_3 = d_{33,m} T_{3m} + \mu_{33}^T H_3, \quad (1b)$$

where $H_3$ and $B_3$ are the externally applied ac magnetic field and magnetic induction in the length direction of the magnetostrictive flake, respectively; $S_{3m}$ and $T_{3m}$ are the strain and stress, respectively; $s_{33}^H$, $d_{33,m}$ and $\mu_{33}^T$ are the elastic compliance coefficient, piezomagnetic coefficient and magnetic permeability, respectively.

The piezoelectric constitutive equations can be expressed as

$$S_{1p} = s_{11}^E T_{1p} + d_{31,p} E_3, \quad (2a)$$

$$D_3 = d_{31,p} T_{1p} + \varepsilon_{33}^T E_3, \quad (2b)$$

where $E_3$ and $D_3$ are the electric field and electric displacement along the thickness direction of the piezoelectric flake, respectively; $T_{1p}$ and $S_{1p}$ are the stress and strain, respectively; $s_{11}^E$, $d_{31,m}$, and $\varepsilon_{33}^T$ are the elastic compliance coefficient, piezoelectric coefficient and dielectric permeability, respectively.

The magnitude of force $F$ applied on the glass slide is

$$F = T_{3m} A_m = \left( \frac{\mu_0 M}{d_{33,m}} + \frac{\mu_0 - \mu_{33}^T}{d_{33,m}} H_3 \right) A_m, \quad (3)$$

where $A_m$ is the cross-sectional area of the magnetostrictive flake, $\mu_0 (=4\pi \times 10^{-7}$ H m$^{-1}$) is the magnetic permeability of free space.

Considering the glass is rigid and the TbFe$_2$, PZT, and glass were perfect bonded. The stress in PZT flakes due to the transfer of $F$ from TbFe$_2$ flake can be expressed as

$$T_{1p} = \frac{1/2 F}{A_p} = \left( \frac{\mu_0 M}{d_{33,m}} + \frac{\mu_0 - \mu_{33}^T}{d_{33,m}} H_3 \right) \frac{A_m}{2 A_p}, \quad (4)$$

where $A_p$ is the cross-sectional area of the piezoelectric flake.

Assuming that the strain in the TbFe$_2$ and PZT flakes is equal, with Eqs. (1), (2), and (4), the ME voltage coefficient can be obtained as

$$\alpha_{E,31} = \frac{dE_3}{dH_3} = \frac{(\mu_0 - \mu_{33}^T)(2 s_{33}^H A_p - s_{11}^E A_m) + 2 d_{33,m}^2 A_p}{2 d_{33,m} d_{31,p} A_p}. \quad (5)$$

Equations (4) and (5) provide an insight into the physics about the transformation of the stress and the enhancement of $\alpha_{E,31}$ in the proposed structure.

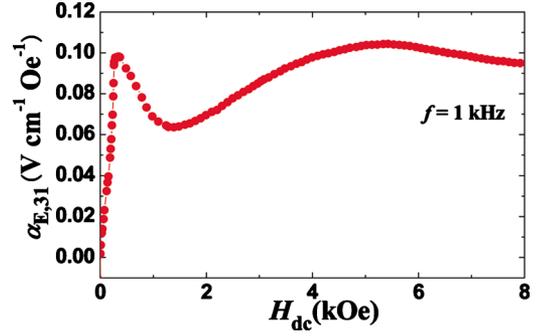

FIG. 2. (Color online) ME voltage coefficient ($\alpha_{E,31}$) as a function of bias magnetic field ($H_{dc}$) at $f=1$ kHz.

Under ferromagnetic and electromechanical resonances, the ME effect is known to be obviously enhanced.[4,16] Hence, the prediction of the resonance frequency is quite important. The resonance frequency of the bending resonance mode is given by[17]

$$f_{r1} = \frac{\pi d}{4\sqrt{3} l^2} \sqrt{\frac{1}{\bar{\rho} \overline{s_{11}}}} \beta_n^2, \quad (6)$$

where $d=1.2$ mm is the total thickness of the PZT flakes, $\beta_n \approx n+1/2$, where $n$ is the order of the bending mode, $\bar{\rho}$ is the average density determined by $\bar{\rho} = v_H \rho_H + v_E \rho_E$, where $v_H$ and $v_E$ are the volume fractions, $\rho_H$ and $\rho_E$ are the density of the TbFe$_2$ and PZT flakes, respectively. The equivalent elastic compliance $\overline{s_{11}}$ is given by

$$\overline{s_{11}} = \frac{s_{11}^E s_{33}^H}{v_E s_{33}^H + v_H s_{11}^E}. \quad (7)$$

The resonance frequency of the planar acoustic resonance mode is given by

$$f_{r2} = \frac{1}{2l} \sqrt{\frac{1}{\bar{\rho} \overline{s_{11}}}}. \quad (8)$$

Using Eqs. (6) and (8), one can predict the first, second, third, and forth bending resonance frequencies to be 12.9, 36.1, 70.7, and 116.8 kHz, and the planar acoustic resonance frequency to be 84.8 kHz. The parameters for the prediction are: $\rho_H = 9 \times 10^3$ kg m$^{-3}$; $\rho_E = 7.45 \times 10^3$ kg m$^{-3}$; $s_{33}^H = 17 \times 10^{-12}$ m$^2$ N$^{-1}$; $s_{11}^E = 13 \times 10^{-12}$ m$^2$ N$^{-1}$.

The dc bias magnetic field ($H_{dc}$) and ac magnetic field ($\delta H$) were applied along the length direction of the flakes. The induced voltage signal $\delta V$ was amplified and measured. The ME voltage coefficient was calculated based on $\alpha_E = \delta V / (t_{PZT} \delta H)$, where $t_{PZT}$ is the total PZT thickness. Figure 2 shows the ME voltage coefficient ($\alpha_{E,31}$) as a function of bias magnetic field ($H_{dc}$) at $f=1$ kHz. The $\alpha_{E,31}$ depends strongly on $H_{dc}$. On increasing $H_{dc}$ from zero, $\alpha_{E,31}$ increases linearly until a maximum value is reached at $H_{dc}=350$ Oe, and then decreases subsequently. With further increase in $H_{dc}$, $\alpha_{E,31}$ reaches a maximum value, and then decreases.

The characteristic in Fig. 2 is related to the variation in the piezomagnetic coupling coefficient $q$ with $H$.[18] The values of the $\alpha_{E,31}$ are proportional to $q \sim \delta \lambda / \delta H$, where $\lambda$ is the magnetostriction. The piezomagnetic coupling coefficient $q$ as a function of applied magnetic field for TbFe$_2$ flake is shown in Fig. 3. The inset shows the dependence of in-plane parallel magnetostriction $\lambda_\parallel$ on the applied magnetic field for TbFe$_2$ flake. It is seen that the variation in $q$ with $H$ is similar to that of $\alpha_{E,31}$ with $H_{dc}$. As $H$ is increased from zero, one observes an increase in $q$ to a maximum at a certain $H$, and then decreases subsequently. With further increase in $H$, $q$ increases until a maximum value is reached, and then decreases subsequently. The behavior of $q$ for TbFe$_2$ flake results in the characteristic observed in Fig. 2.

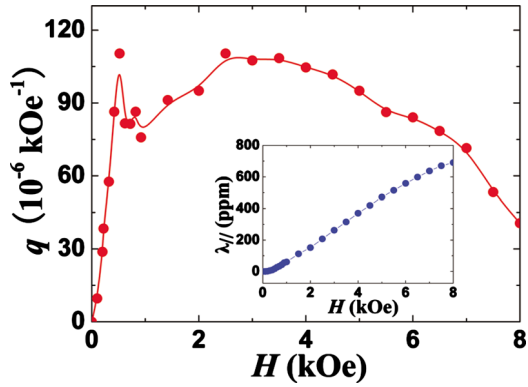

FIG. 3. (Color online) Piezomagnetic coupling coefficient ($q$) as a function of applied magnetic field for the TbFe$_2$ flake. The inset shows the dependence of in-plane parallel magnetostriction $\lambda_\parallel$ on the applied magnetic field for the TbFe$_2$ flake.

Figure 4 shows the ME voltage coefficient ($\alpha_{E,31}$) as a function of frequency ($f$) at $H_{dc}$=350 Oe. There are five resonance peaks for $1 < f < 150$ kHz and a large ME voltage coefficient as high as $\alpha_{E,31}$=10.5 V cm$^{-1}$ Oe$^{-1}$ can be obtained at resonance frequency of $f$=85.5 kHz, which is larger than that of conventional layered TbFe$_2$/PZT composites.[7] One can easily identify that the resonance peaks at 12.0 kHz, 36.2 kHz, 69.5 kHz, 116.5 kHz are ascribed to the first, second, third, and forth bending resonance modes, respectively, while the resonance peak at 85.5 kHz is attributed to the first planar acoustic resonance mode. The experimental results are in good agreement with the predicted ones mentioned above. It need be noted that the maximum value of $\alpha_{E,31}$=9.9 V cm$^{-1}$ Oe$^{-1}$ at 69.5 kHz for the third bending resonance was also observed.

In summary, we reported a large ME effect in a structure mechanically mediated by a stretching force and made up of TbFe$_2$, PZT, and nonmagnetic flakes. The significant characteristic of the proposed structure is the large ME coupling achieved by end bonding instead of interface bonding. The resonance frequencies have been theoretically and experimentally studied. A maximum values of $\alpha_{E,31}$=10.5 V cm$^{-1}$ Oe$^{-1}$ was obtained at first planar acoustic resonance frequency of $f$=85.5 kHz, which makes this proposed structure suitable for applications in sensors, actuators and transducers.

This work is supported by the Natural Science Foundation of Jiangsu Province of China (Grant No. BK2010505) and the Funding of Jiangsu Innovation Program for Graduate Education (Grant No. CX10B_099Z). D. A. Pan would like to acknowledge support from the National Natural Science Foundation of China under Grant No. 50802008.

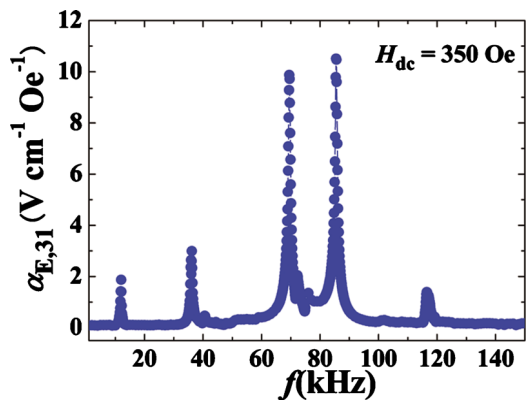

FIG. 4. (Color online) ME voltage coefficient ($\alpha_{E,31}$) as a function of frequency ($f$) at $H_{dc}$=350 Oe.